\documentclass[fleqn,usenatbib]{mnras}
\usepackage{newtxtext,newtxmath}
\usepackage[T1]{fontenc}
\usepackage{ae,aecompl}
\usepackage{array}

\usepackage{graphicx}	
\usepackage{amsmath}	
\usepackage{amssymb}	

\usepackage{textcomp}
\usepackage{color, colortbl}

\definecolor{Gray}{gray}{0.9}
\definecolor{LightCyan}{rgb}{0.88,1,1}

\usepackage{natbib}
\usepackage{amsmath}
\usepackage{graphicx}
\usepackage[colorinlistoftodos]{todonotes}
\usepackage{xcolor}
\usepackage{multirow}
\usepackage{todonotes}
\usepackage{float}
\usepackage{booktabs}
\usepackage{subfig}
\usepackage{lineno}

\newcommand{\bs}[1]{\boldsymbol{#1}}
\newcommand{\mc}[1]{\mathcal{#1}}

\defcitealias{Hosenie_2019}{H19}

\title[Imbalanced Learning for Variable Star Classification]{Imbalance Learning for Variable Star Classification}

\author[Z. Hosenie et al.]{
Zafiirah Hosenie,$^{1}$\thanks{E-mail: zafiirah.hosenie@gmail.com}
Robert Lyon,$^{1,2}$
Benjamin Stappers,$^{1}$
Arrykrishna Mootoovaloo,$^{3}$ 
\newauthor 
and Vanessa McBride$^{4,5,6}$
\\
$^{1}$Jodrell Bank Centre for Astrophysics, Department of Physics and Astronomy, The University of Manchester, Manchester M13 9PL, UK.\\
$^{2}$Department of Computer Science, Edge Hill University, Ormskirk Lancashire L39 4QP, UK.\\
$^{3}$Imperial Centre for Inference and Cosmology (ICIC), Imperial College, Blackett Laboratory, Prince Consort Road, London SW7 2AZ, UK.\\
$^{4}$Department of Astronomy, University of Cape Town, Private Bag X3, Rondebosch, 7701, South Africa.\\
$^{5}$South African Astronomical Observatory, PO Box 9, Observatory, 7935, South Africa.\\
$^{6}$IAU Office of Astronomy for Development, Cape Town, Observatory, 7935, South Africa.\\
}

\date{Accepted 2020 February 27. Received 2020 February 27; in original form 2019 October 23}

\pubyear{2020}

\begin{document}
\label{firstpage}
\pagerange{\pageref{firstpage}--\pageref{lastpage}}
\maketitle

\begin{abstract}
The accurate automated classification of variable stars into their respective sub-types is difficult. Machine learning based solutions often fall foul of the imbalanced learning problem, which causes poor generalisation performance in practice, especially on rare variable star sub-types. In previous work, we attempted to overcome such deficiencies via the development of a hierarchical machine learning classifier. This `algorithm-level' approach to tackling imbalance, yielded promising results on Catalina Real-Time Survey (CRTS) data, outperforming the binary and multi-class classification schemes previously applied in this area. In this work, we attempt to further improve hierarchical classification performance by applying `data-level' approaches to directly augment the training data so that they better describe under-represented classes. We apply and report results for three data augmentation methods in particular: \textit{R}andomly \textit{A}ugmented \textit{S}ampled \textit{L}ight curves from magnitude \textit{E}rror (\texttt{RASLE}), augmenting light curves with Gaussian Process modelling (\texttt{GpFit}) and the Synthetic Minority Over-sampling Technique (\texttt{SMOTE}). When combining the `algorithm-level' (i.e. the hierarchical scheme) together with the `data-level' approach, we further improve variable star classification accuracy by 1-4\%. We found that a higher classification rate is obtained when using \texttt{GpFit} in the hierarchical model. Further improvement of the metric scores requires a better standard set of correctly identified variable stars and, perhaps enhanced features are needed.
\end{abstract}

\begin{keywords}
stars: variables- general -- methods: data analysis - Astronomical instrumentation, methods, and techniques.
\end{keywords}



\section{Introduction}

Astronomy is now in an era dominated by an explosion of big data, produced with current and future surveys, such as OGLE \citep{udalski2008optical, udalski2015ogle}, CRTS \citep{Drake_2017} and Kepler \citep{koch2010kepler} among others, thus, relying solely on visual inspection for classification is becoming impractical. Therefore, automatic classification pipelines are required to categorize an unprecedented amount of variable star light curves into known or unknown classes for various astrophysical applications. Accordingly, machine learning has heavily been studied to solve classification problems, for instance, uncovering aberrant phenomena encountered in observations, also known as unsupervised anomaly detection \citep{chen2018anomaly, zong2018deep} and automatic classification of variable stars \citep{kim2016package, benavente2017automatic, Mahabal2017, Narayan_2018, pashchenko2018machine, tsang2019deep, zorich2020streaming}.

\begin{figure*}
\centering
\includegraphics[width=0.75\textwidth]{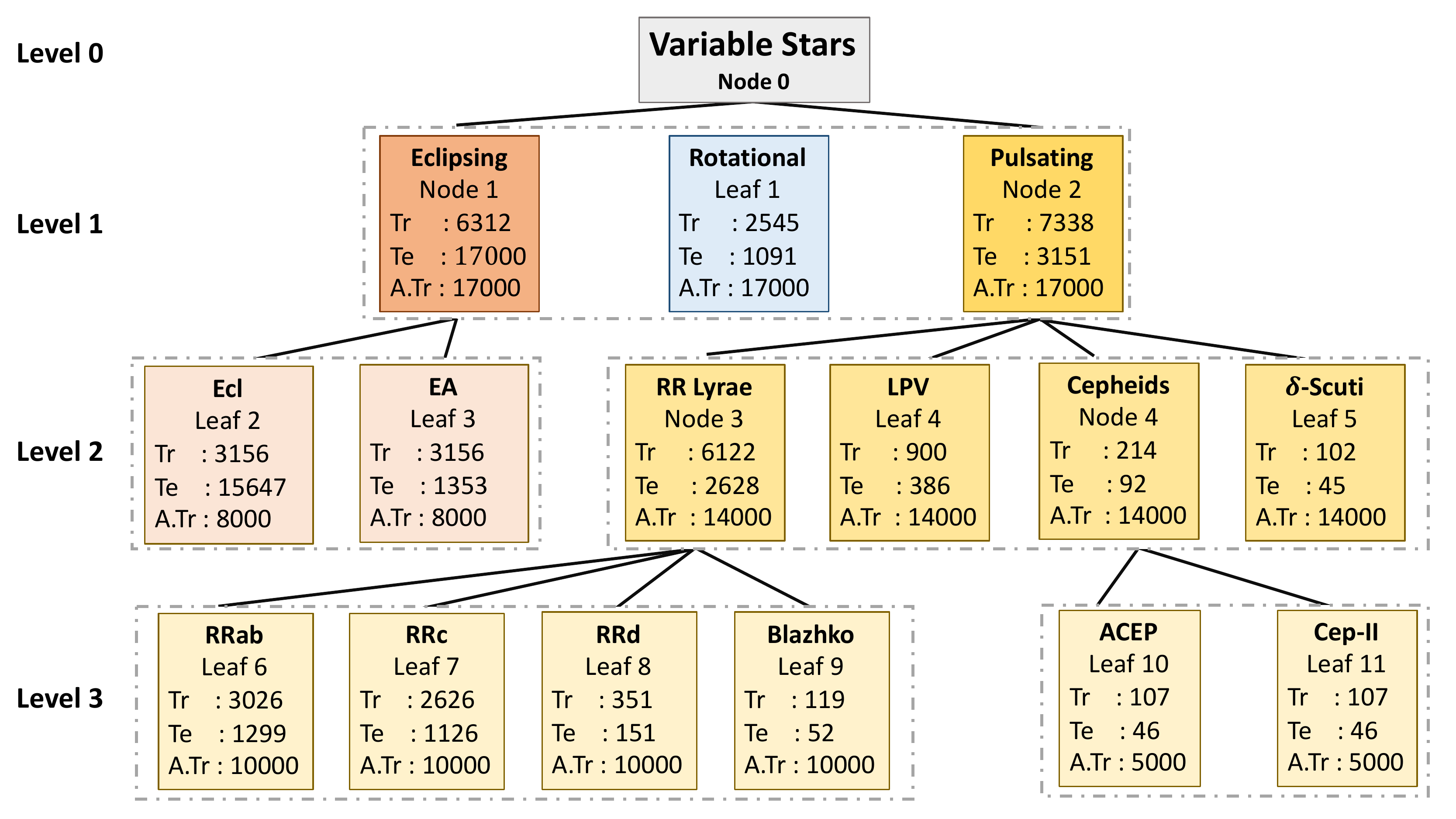}
\caption{Hierarchical Tree classification with automated light curves augmentation for CRTS Data. The number of training examples (real LCs) is represented by \textit{Tr}, the number of training examples after augmentation (both real and synthetic LCs) is represented by \textit{A.Tr} and the number of test examples (real LCs) is represented by \textit{Te}. At level 1, the real LCs in the training set are augmented and the dotted square represents a trained model (RF/XGBoost classifier). During testing phase, the classified examples in the test set move down the hierarchy at level 2. Afterwards, real LCs in the training set in level 1 moves to their respective branches at level 2. The real LCs are augmented and features are extracted. This process is repeated until it reaches all leaves in the hierarchy.}
\label{fig:Class-distribution-of-full-dataset}
\end{figure*}

However, a major issue that impedes the successful automated classification of astronomical data is known as the imbalanced learning problem. This occurs when we wish to organise data into distinct groups known as ``classes'', using examples to guide a process known as ``classification". When there is a large distributional difference between the number of examples belonging to each class, minority  and majority classes form. When the imbalance between the minority and majority classes is large, problems can arise when attempting to build standard machine learning classification algorithms, ultimately resulting in poor categorisation performance. This happens as such algorithms are usually optimised to achieve maximum accuracy. However this is trivially achievable in imbalanced datasets by always assigning the majority class label when making predictions. This leads to biased classifiers that obtain high predictive accuracy for majority class, but poor predictive accuracy for minority classes, which are more often than not, the focus of our interest.

Imbalanced learning problems occur in many domains, for instance in fraudulent phone call identification (few calls are fraudulent, \citep{Fawcett_1996}), or text classification (in cases where there is either more positive or more negative sentiments). In astronomy, this issue becomes acute given that datasets must often be searched for rare or unusual phenomena which may not be accurately defined in advance. This problem impacts the classification of variable stars in particular, as some types of variable star are uncommon, making it difficult to build systems to be able to recognise them. In astronomy, several works have tried to address the problem of class imbalance to date (\citealt{hoyle2015data, Lochner_2016, Narayan_2018, Revsbech_2018, agarwal2019towards}).

There are two approaches for dealing with class imbalance problems \citep{he2008learning}. The first are generally known as `algorithm level' approaches. These seek to modify classification algorithms directly, to better accommodate imbalanced class distributions. This can involve, for example, adapting the learning function at the heart of the algorithm to favour metrics other than accuracy during training and also applying hyperparameter tuning while training the algorithm (See \S \ref{Training with Bayesian Optimization}). Algorithm level approaches make an implicit assumption - that the data is sufficiently descriptive and statistically characteristic of the classes under consideration, and changes to the algorithm alone will enable this data to yield good classification performance.

Alternatively, `data level' approaches seek to modify the data given to a classification algorithm, with the aim of improving classification performance. Data level approaches can be as simple as balancing training data artificially via an appropriate sampling method, or as complex as generating artificial samples to balance the training set. Data level approaches assume that classification algorithms will be capable of separating the classes under consideration, given appropriate training data. Hybrid approaches mix the two techniques when faced with difficult problems. For instance, in some cases modifying an algorithm will not produce the improvement expected, if the classification problem at hand exhibits excessive class overlap, disjuncts, or is affected by small sample sizes (i.e. some classes are genuinely rare). Whilst in some cases trying to balance training sets will not work if the information content of the training samples is too low to allow a classifier to delineate effective class boundaries.

In previous work, we attempted to develop a variable star classifier together with various techniques of feature selection and feature importance, and ran into the imbalanced learning problem. To overcome this, we attempted to modify the algorithms used for classification, and ultimately proposed a successful hierarchical classification system. We compared the hierarchical system (using 7 features) with the \texttt{UPSILON} package \citep{kim2016package} (using 16 features). Whilst hierarchical system was effective, recall on minority classes could be stubbornly low relative to majority classes. In other domains such problems are overcome by balancing the training distribution directly. This approach implies the minority class is sufficiently described in the training data to solve the imbalance, and further that the classifier used is sensitive to the class size. We believe this to be the case, thus we proceed similarly. We present a hybrid approach to overcoming imbalances, which represents a principled and pragmatic approach to this problem. Thus in this work, we improve the \citet[][hereafter H19]{Hosenie_2019} classification scheme by adding a sufficient amount of data, such that each class has an equal amount of training examples. This can be achieved by simulating more data or gathering more real data (which is often difficult). 

Balancing training sets directly can be difficult. Fortunately, techniques such as Synthetic Minority Over-sampling Technique (\texttt{SMOTE}, \citealt{Chawla}), random values drawn from the Gaussian distribution \citep{peterson1998uncertainties} and Gaussian Processes (GPs, \citealt{Rasmussen_2005}) modelling (\texttt{GpFit}) can simplify the problem to a large extent by simulating lightcurves. GPs have been used in several works to synthetically augment biased supernova training sets \citep{Lochner_2016,Narayan_2018,Revsbech_2018}, variable stars \citep{faraway2016modeling, castro2018uncertain, martinez2018high} and lightcurve detrending \citep{aigrain2016k2sc}.

In this work, we are concerned only with periodic variable star classification and we present GPs for augmenting periodic variable star data using folded light curves. Second, we propose a new method, \textit{R}andomly \textit{A}ugmented \textit{S}ampled \textit{L}ight curves from magnitude \textit{E}rror (\texttt{RASLE}\footnote{After the preparation of this manuscript, we learnt that another team \citet{gabruseva2019photometric}, has come up with a similar method independently.}), to periodic variable star data for the first time, which synthetically augments the training set by sampling from the magnitude errors. We then compare the three data augmentation methods (\texttt{SMOTE}, \texttt{GpFit} $\&$ \texttt{RASLE}) and their utility for improving variable star classification, trained with either a Random Forest (RF, \citealp{Breiman_2001}) classifier or eXtreme Gradient Boosting (XGBoost, \citealp{Chen_2016}) classifier. Finally, we incorporate a Bayesian Optimisation approach to find the best hyper-parameters for the RF and XGBoost in the hierarchical classification (HC) scheme. We achieve an improvement of 1-4 percent in terms of balanced-accuracy and G-mean scores at all levels in the HC, compared to the results of \citetalias{Hosenie_2019}.  

The structure of the paper is as follows. In \S \ref{sec:CRTS_DATA}, we describe the data set used in our analysis; while in \S \ref{subsec:data_augmentation}, the three data augmentation algorithms used, are explored. In \S \ref{sec:Descrition_methodology}, we provide a description of the various stages in the hierarchical classification pipeline; in \S \ref{sec:Analysis_and_Results} we present the classification results and finally, we conclude in \S \ref{sec:Conclusion_2}.

\section{DATA}\label{sec:CRTS_DATA}
The Catalina Real-Time Transient Survey (CRTS, \citealt{Drake_2017}) has produced a catalogue of  periodic variable stars from 6 years of optical photometry from the Siding Spring Survey (SSS). We consider only 11 classes from the CSDR2\footnote{\href{http://nesssi.cacr.caltech.edu/DataRelease/VarcatS.html}{Catalina Surveys Data Release 2}} dataset as presented in Table \ref{tab:class_distribution_of_CRTS} for our analysis. From Table \ref{tab:class_distribution_of_CRTS}, we observe that the data is heavily imbalanced. Thus to simplify our experimentation, we reduced the size of the largest class (Ecl) via random under-sampling. We down-sample this class to 4509 (this makes the number of Ecl examples comparable to the next biggest class, EA) and the remaining Ecl light curves (LCs) are then used for prediction. This is why the number of samples available for testing exceeds those for training as shown in Fig \ref{fig:Class-distribution-of-full-dataset}.

\renewcommand{\arraystretch}{1.1}
\begin{table}
\begin{minipage}{75mm}

\caption{Sample size of classes in CRTS data. The class distribution is extremely imbalanced, such as Ecl are over-represented. \label{tab:class_distribution_of_CRTS}}
\noindent \begin{centering}
\begin{tabular}{p{5.5cm}>{\centering}p{0.5cm}||>{\centering}p{0.5cm}}
\hline 
\textbf{Types of variable stars} & \multicolumn{2}{c}{\textbf{$\textrm{N}_{\textrm{Objects}}$}}\tabularnewline
\hline 
\hline 
RRab (fundamental mode) & \multicolumn{2}{c}{4325}\tabularnewline
RRc (first overtone mode) & \multicolumn{2}{c}{3752}\tabularnewline
RRd (multimode) & \multicolumn{2}{c}{502}\tabularnewline
Blazhko (long-term modulation) & \multicolumn{2}{c}{171}\tabularnewline
Contact \& Semi-Detached Binary: Ecl & \multicolumn{2}{c}{18803}\tabularnewline
Detached Binary: EA & \multicolumn{2}{c}{4509}\tabularnewline
Rotational: Rot & \multicolumn{2}{c}{3636}\tabularnewline
Long Period Variable: LPV & \multicolumn{2}{c}{1286}\tabularnewline
$\delta$-Scuti & \multicolumn{2}{c}{147}\tabularnewline
Anomalous Cepheids: ACEP & \multicolumn{2}{c}{153}\tabularnewline
Type-II Cepheids: Cep-II & \multicolumn{2}{c}{153}\tabularnewline
\hline 
\end{tabular}
\par\end{centering}
\end{minipage}
\end{table}

\section{Data Augmentation}\label{subsec:data_augmentation}
While the under-sampling methods (i.e. downsample Ecl and developing the hierarchical system) help to address some of the class imbalance issues, they are themselves insufficient, as minority class performance was not good enough for our purposes. We therefore provide three ways to over-sample the data, a form of data augmentation necessary as some of the classes still outnumber other classes (see \textit{Tr} values in Fig \ref{fig:Class-distribution-of-full-dataset}). We augment the data via the generation of artificial data, in order to increase the number of training samples by generating similar but not identical examples. In principal the more data we have, the better our ML models will be as this technique helps to reduce overfitting. In this work, we consider three methods of augmentation, (i) \texttt{SMOTE}, (ii) \texttt{RASLE}, and (iii) \texttt{GpFit}.

\subsection{Synthetic Minority Over-sampling Technique}
The Synthetic Minority Over-sampling Technique (\texttt{SMOTE}) inserts artificially generated minority class examples into a dataset, by operating in \textquotedblleft \textit{feature space}\textquotedblright{} rather than \textquotedblleft \textit{data space}\textquotedblright . This technique helps to balance the overall class distribution. The standard implementation of \texttt{SMOTE} utilizes $k-$nearest neighbours \citep{Buturovic_1993} to group similar class objects and to determine which class categories are in the minority class and need over-sampling. To generate a new synthetic example, the $k-$nearest neighbours method is further used by first selecting an example in the minority class. The collection of feature values describing this example, it's feature vector, is then combined with the feature vectors of one of it's $k$ nearest neighbours chosen at random. The difference between the vectors of these two examples is computed and subsequently multiplied by a random number drawn between 0 and 1. This produces an entirely new synthetic feature vector. This process is repeated until enough synthetic examples have been created. Finally, the new augmented training set is comprised of both the synthetic examples and the real minority examples. In our pipeline, we utilize the `\texttt{regular-SMOTE}' algorithm from the imbalanced-learn\footnote{\href{https://imbalanced learn.readthedocs.io/en/stable/index.html}{https://imbalanced-learn.readthedocs.io/en/stable/index.html}} \citep{LeMaitre_2017} package.

\subsection{\textit{R}andomly \textit{A}ugmented \textit{S}ampled \textit{L}ight curves from magnitude \textit{E}rrors}
The artificial examples generated by standard \texttt{SMOTE}, may not truly represent data recorded during observations. One way around this is to generate artificial samples from existing data points in a more scientifically valid way. That is we randomly sample a selection of rare class examples, take their primary characteristics, and generate new examples from them by perturbing them in a principled way. We do this using the \textit{R}andomly \textit{A}ugmented \textit{S}ampled \textit{L}ight curves from magnitude \textit{E}rrors (\texttt{RASLE}) method.

\begin{figure}
\centering
\includegraphics[width=0.5\textwidth]{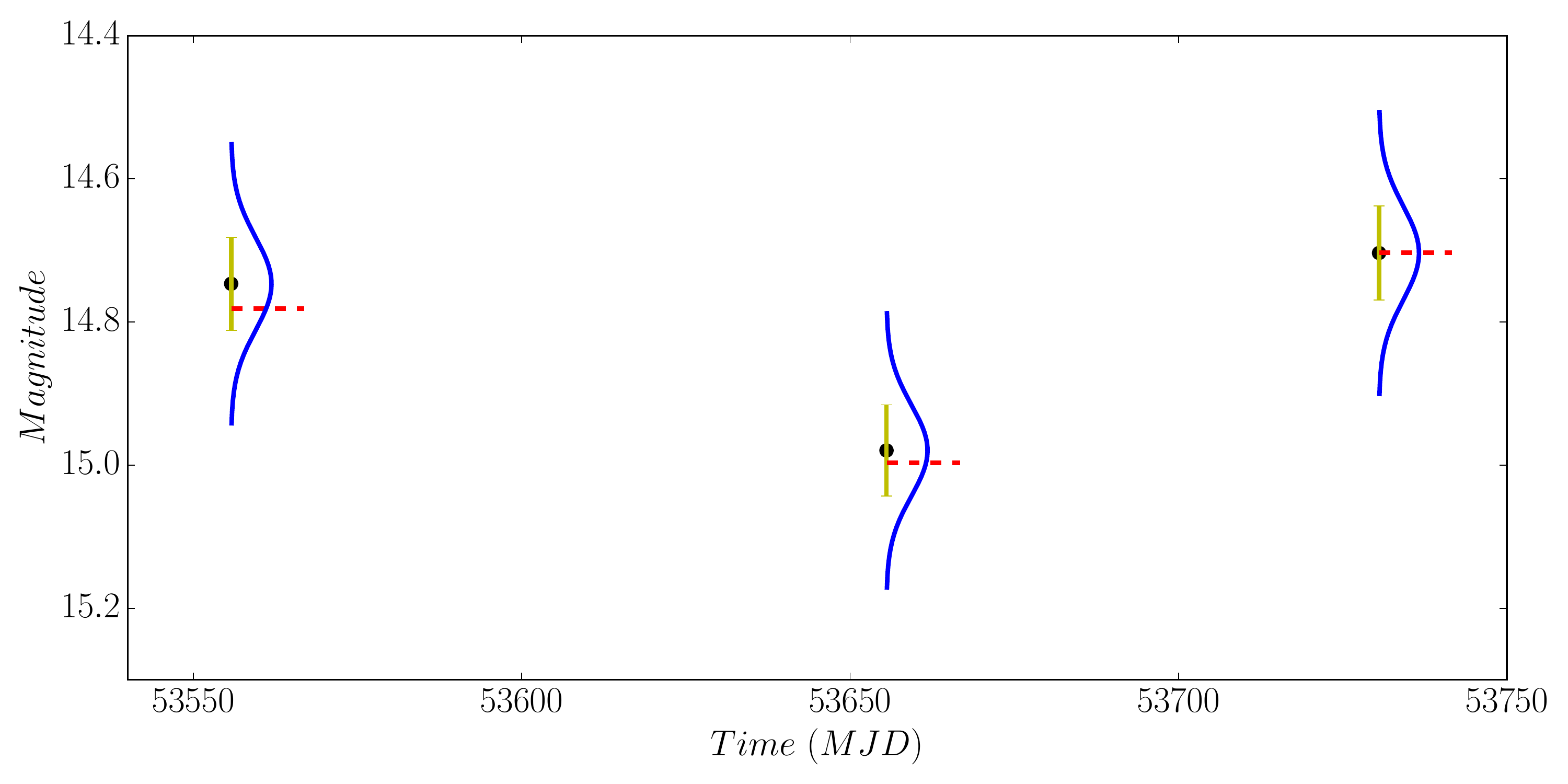}
\caption{Generating new light curves by random sampling from a normal distribution. The true magnitude along with its error bars is shown in black and yellow. We assume a normal distribution with mean equal to the true magnitude and with sigma equal to the error in magnitude. We randomly draw one sample (red-dashed line) from each normal distribution to produce a completely new light curve.\label{fig:RSLC}}
\end{figure}

The application of \texttt{RASLE} is employed on unfolded-LCs, that is, each variable star is described by its time, magnitude and error in magnitude. Using this information, we generate new light curves in the following way. Let us consider a probability distribution which can be concisely represented by a normal distribution. The probability distribution function (\textit{pdf}) can be interpreted as going over the magnitude space vertically with the horizontal axis showing the probability that some value will occur. To construct the \textit{pdf}, we make an assumption that the magnitude follows a normal distribution with mean, $\mu$, to be equal to the true magnitude and the standard deviation, $\sigma$, to be equal to the error in magnitude. For each data point at a specific time, we sample a single magnitude from the \textit{pdf}. Each sampled magnitude is assigned the same time as in the original data. Fig \ref{fig:RSLC} shows an example of a light curve with the magnitude and error bars drawn for three specific times. The \textit{pdf} of the magnitude is shown in blue and one magnitude is sampled randomly from the \textit{pdf} shown in dotted red lines. The generated light curve is given the new (random) sampled magnitude with the same time value as in the original data.

\begin{figure*}
\centering
\includegraphics[width=\textwidth]{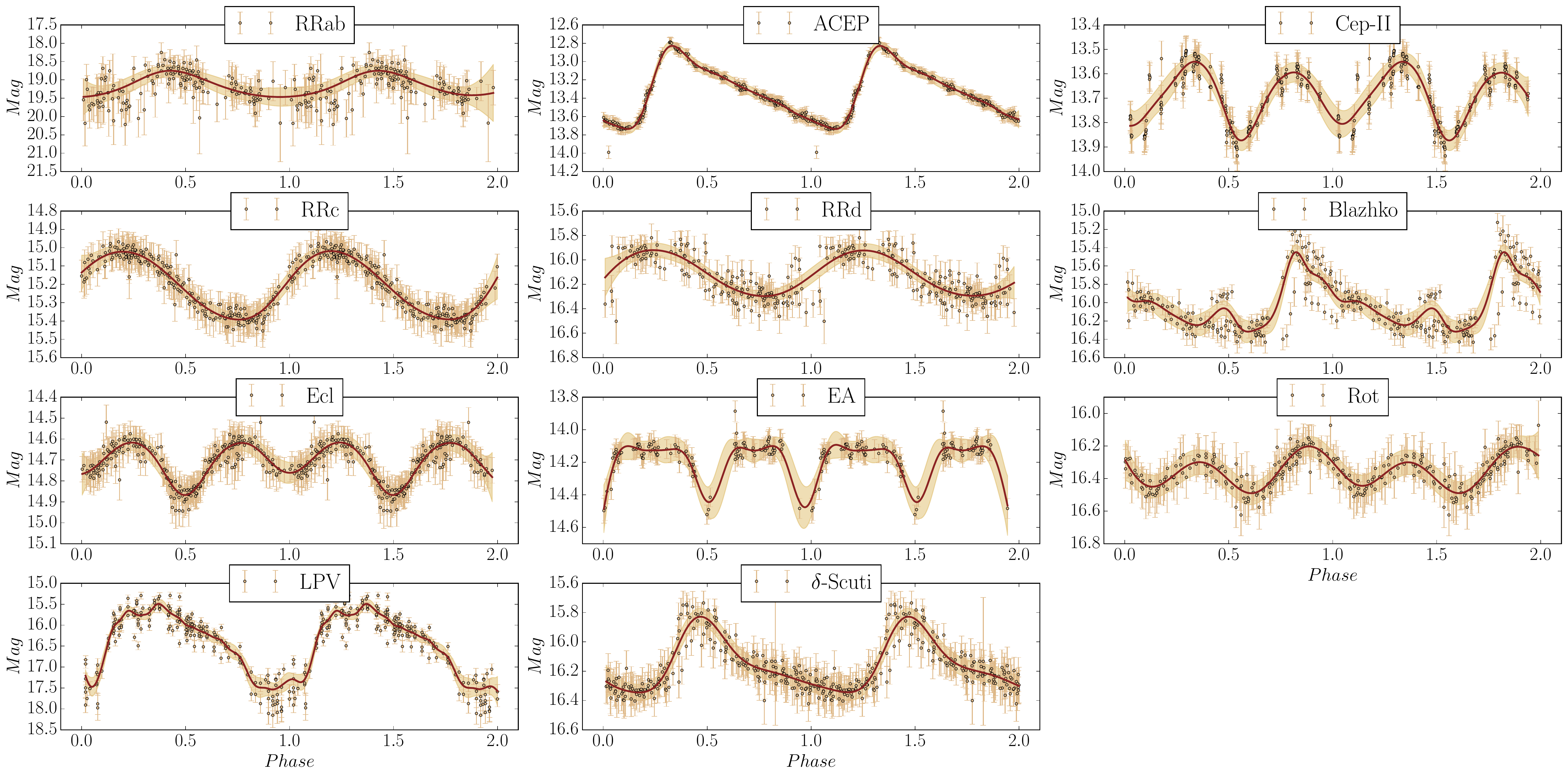}
\includegraphics[width=0.90\textwidth]{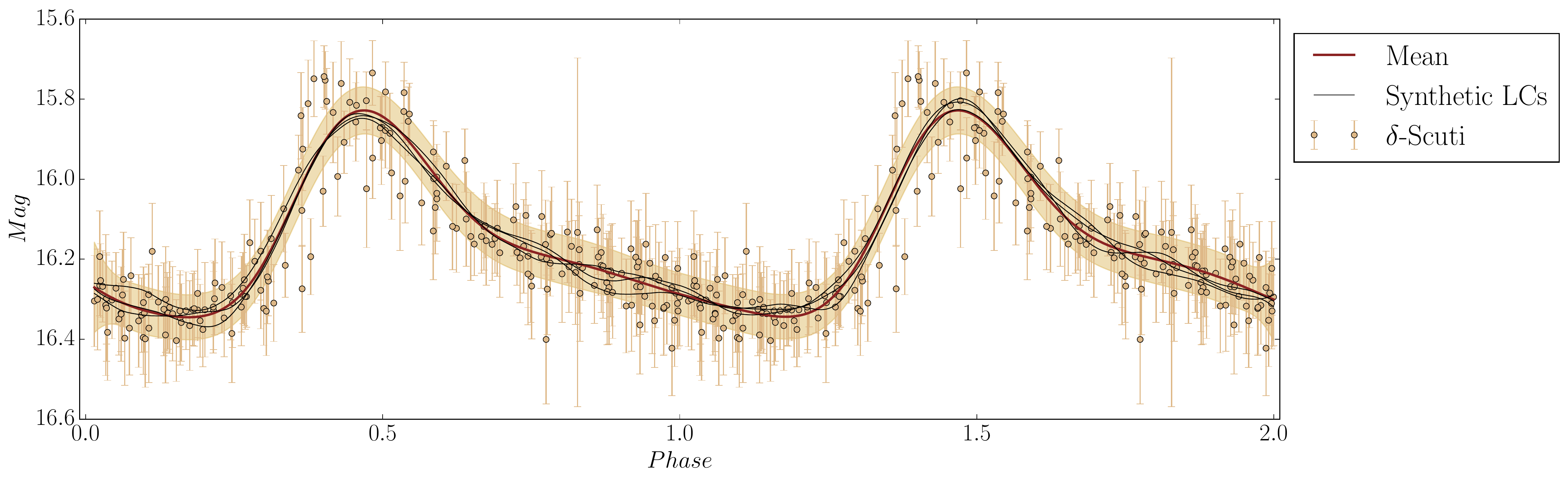}
\caption{Gaussian Processes offer a flexible approach to produce a smooth model of periodic light curves reported in magnitudes as a function of phase. This is demonstrated with model fits for each example of variable stars considered in the CRTS dataset. The data points are illustrated in black-rounded dots along with the error bars. The mean of the GP fit is shown in brown with three standard deviation away from the mean, shown in shaded pale brown. In the bottom panel, the black lines represent three randomly drawn samples from the GP fit. These randomly sampled light curves, also known as synthetic LCs together with real LCs, are used in the training set.\label{fig:GP-Fit}}
\end{figure*}

\subsection{Modelling Light Curves with Gaussian Process}
An ideal case for data augmentation is to use a well defined model of the classes under consideration to create synthetic data. However, there is no available model valid for all the different variable stars considered. We therefore build a model describing variable stars using Gaussian Processes (GPs, \citealt{Rasmussen_2005}) applied to CRTS data. We then use this model to generate artificial light curves, allowing us to augment our training data through the addition of new examples in a principled way, using the distributions of existing data to create them.

 A GP is a distribution over functions. It is defined by a mean $\mu(t)$ and a covariance (kernel) function $c(t,t')$, and is given as 

\begin{equation}
f(t)\,\sim\,\textrm{GP}\left(\mu(t),c(t,t')\right).
\end{equation}

\noindent When the function $f$ is computed at points $t$, the marginal distribution follows a multivariate normal distribution \citep{Rasmussen_2005}. The kernel function, $c$, takes two inputs and shows the similarity between them. When evaluating Bayesian inference, having the set of known function values for the training sets $\boldsymbol{f_{x}}$, and the set of known function values for the test sets $\boldsymbol{f_{y}}$, are normally distributed and is given as follows:
 
\begin{equation}
\left[\begin{array}{c}
\bs{f_{x}}\\
\bs{f_{y}}
\end{array}\right]\,=\,\mathcal{N}\left(\begin{array}{ccc}
\left[\begin{array}{c}
\bs{\mu}_{f_{x}}\\
\bs{\mu}_{f_{y}}
\end{array}\right] , \left[\begin{array}{cc}
\bs{C}_{f_{x}f_{x}} & \bs{C}_{f_{x}f_{y}}\\
\bs{C}_{f_{x}f_{y}} & \bs{C}_{f_{y}f_{y}}
\end{array}\right]\end{array}\right),
\end{equation}

\noindent where the means of the training and test set are denoted by $\bs{\mu}_{f_{x}}$ and $\bs{\mu}_{f_{y}}$ respectively and likewise $\bs{C}_{f_{x}f_{x}}$, $\bs{C}_{f_{y}f_{y}}$, $\bs{C}_{f_{x}f_{y}}$ represent the training, test and train-test covariances/kernels. The conditional distribution, $\bs{f_{x}}\mid \bs{f_{y}}=\mc{P}$ is given by

\begin{equation}\label{eq:conditional_distribution}
\mc{P}\sim\mathcal{N}\left(\bs{C}_{f_{x}f_{y}}\bs{C}_{f_{y}f_{y}}^{-1}\left(\bs{f_{y}}-\bs{\mu}_{y}\right)
+\bs{\mu}_{f_{x}},\bs{C}_{f_{x}f_{x}}-\bs{C}_{f_{x}f_{y}}\bs{C}_{f_{y}f_{y}}^{-1}\bs{C}_{f_{x}f_{y}}^\intercal\right).
\end{equation}

\noindent For a specific set of testing samples, Eq \ref{eq:conditional_distribution}, represents the posterior distribution. For a set of training examples $\mathcal{D}$, the posterior distribution is described by \citep{Rasmussen_2005}

\begin{equation}
\begin{split}
f_{y}\mid\mathcal{D}\,&\sim\,\textrm{GP\ensuremath{\left(\mu_{\mathcal{D}},c_{\mathcal{D}}\right)}},\\
\mu_{\mathcal{D}}(t)\,&=\,\mu(t)+\bs{c}_{T_{s}t}^{\intercal}\bs{C}^{-1}\left(f_{y}-\mu\right),\\
c_{\mathcal{D}}\left(t,t'\right)\,&=\,c\left(t,t'\right)-\bs{c}_{T_{s}t}^{\intercal}\boldsymbol{C}^{-1}\boldsymbol{c}_{T_{s}t'}^{\intercal},
\end{split}
\end{equation}

\noindent where the covariance vector between every training sample, $T_{s}$ and $t$ is $\bs{c}_{T_{s}t}\,=\,c(T_{s},t)$. The choice of the covariance function is established, based on the knowledge of the domain. In our case, we want to model light curves, so we require a kernel that can demonstrate both small fluctuations and smooth variations. Given the different characteristics of the various stars, an appropriate choice of the kernel in this work is the Matern 5/2 kernel given by,
\begin{equation}
C_{\textrm{Matern52}}(\Upsilon)\,=\,\left(1+\frac{\sqrt{5}\Upsilon}{\ell}+
 \frac{5\Upsilon^{2}}{3\ell}\right)\textrm{exp}\left(-\frac{\sqrt{5}\Upsilon}{\ell}\right),
\end{equation}

\noindent where $\Upsilon$ and $\ell$ are the kernel hyperparameters, that is, $\Upsilon$ controls the degree of smoothness and $\ell$ is the characteristic length scale. We employ the GP regression using \texttt{George} \citep{Ambikasaran_2014} with kernel hyper-parameters randomly initialised. Using our data and these randomly initialised hyper-parameters, the negative log likelihood is calculated. Afterwards, these hyper-parameters for the kernel are optimised (i.e., finding the best values for these parameters) using the Limited memory Broyden-Fletcher- Goldfarb-Shanno (L-BFGS, \citealt{Fletcher_1987}) optimization algorithm by minimizing the negative log likelihood.

The kernel with the optimized parameters is then used to fit the GP from which we sample synthetic light curves to augment our training set. Before fitting a GP to our data, we first convert the LCs from time distribution to phase distribution (folded-light curves) where the data is at the detected period for each variable star. We then randomly sampled synthetic LCs from the GP model to form the augmented training set. We show an example of \texttt{GpFit} on the folded-LCs for the different variable stars in Fig \ref{fig:GP-Fit} and the bottom plot illustrates 3 synthetic LCs randomly drawn from \texttt{GpFit}. We then unfolded the phases back into time space and used those synthetic LCs together with the original LCs as the training set.

\section{METHOD DESCRIPTION}\label{sec:Descrition_methodology}
Drawing heavily from \citetalias{Hosenie_2019}, we outline the general approach used to classify variable stars. In this study, we use RF and XGBoost classifiers. We use these classifiers for two reasons. Firstly, to ensure that results presented here are comparable with previous work. Secondly, because they have proven to be robust against the issues associated with class imbalance (\citealt{chen2004using, wang2019imbalance}). We then provide an overview of the HC scheme, together with the various stages we adopt to build the ML pipeline. Similar to \citetalias{Hosenie_2019}, we pre-processed the lightcurves by applying a sigma-clipping method prior to any analysis.

\begin{figure}
\centering
\includegraphics[width=0.5\textwidth]{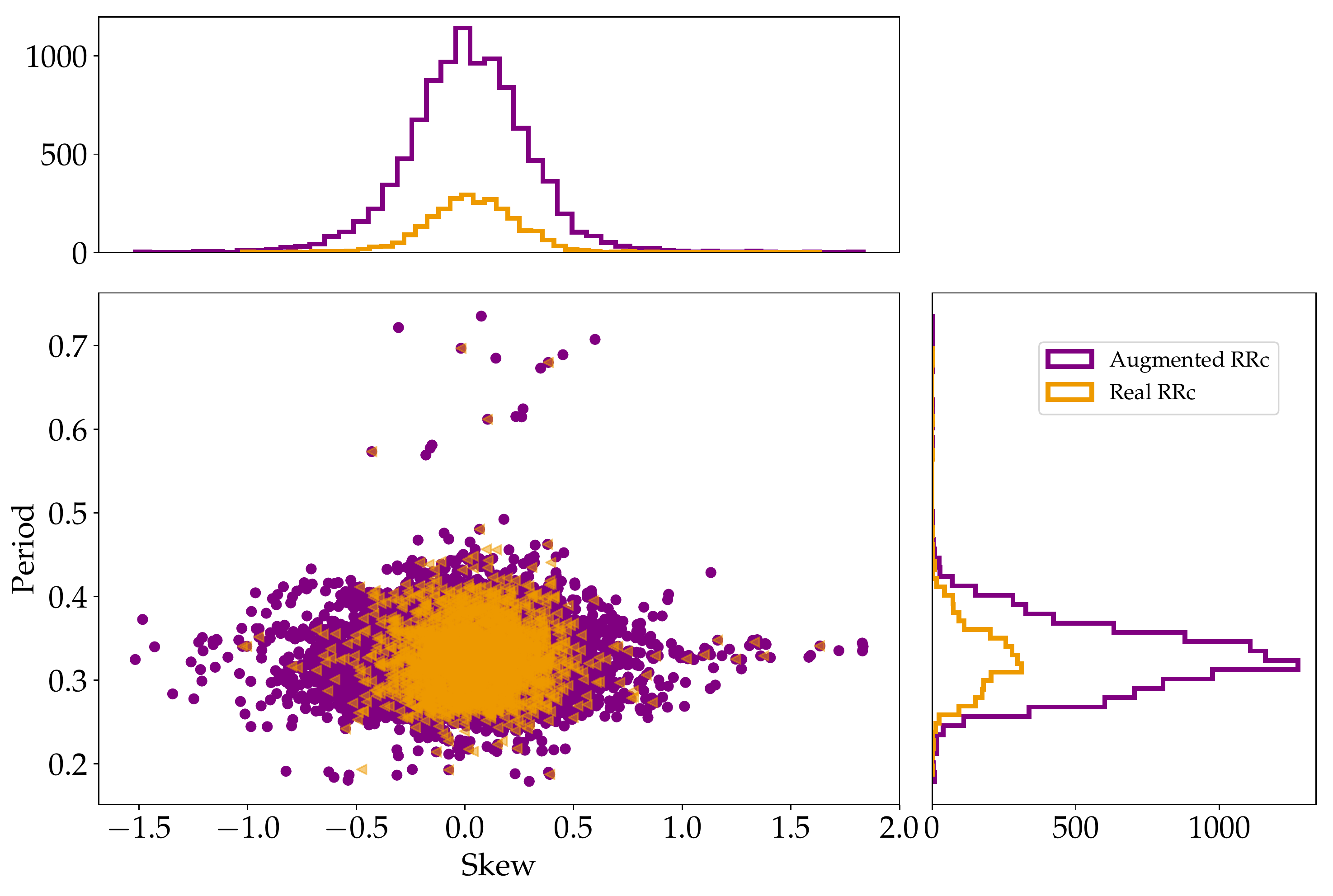}
\caption{Period versus Skew distribution for real and synthetic LCs generated using \texttt{GpFit}. \label{fig:real_vs_aug}}
\end{figure}

\subsection{Stage 1: Hierarchical Tree Classifiers}\label{subsec:HC_Scheme}
\citetalias{Hosenie_2019}'s HC uses the astrophysical properties of the various sources to construct a tree-based structure to represent the different classes (Fig \ref{fig:Class-distribution-of-full-dataset}). Each node/leaf represents a class - identified by the label inside the node/leaf - and the edges represent the relationship between the super-class and sub-class. For the HC, we use XGBoost and RF and then report the one that provides the best classification performance.
XGBoost is a boosting algorithm and is a tree-based model which became popular since its inception in the ML community in 2016. XGBoost works in the same way as Gradient Boosting Decision Tree (GBDT, \citealp{Friedman_2001}). GBDT is an ensemble classification system that iteratively adds simple decision tree classifiers. The first classifier of the ensemble system is trained on the data, while the successive classifiers are trained on the errors of the predecessor classifiers. Unlike, in GBDT, XGBoost parallelizes this process/task and gives a substantial boost in speed. In addition, this classifier controls overfitting by using the regularization techniques, L1-norm \citep{tibshirani1996regression} and L2-norm \citep{ng2004feature}. While a RF is simply an addition of decision trees that aggregate tree decisions. In astronomy, XGBoost has recently been used by \citet{Mirabal_2016} who implemented this classifier for unknown point source classification in the Fermi-LAT catalog and for the separation of pulsar signals from noise \citep{Bethapudi_2018}. In addition, XGBoost has also been applied for variable star classification \citep{sesar2017machine, pashchenko2018machine,kgoadi2019general}.

\begin{figure}
\centering
\includegraphics[width=0.45\textwidth]{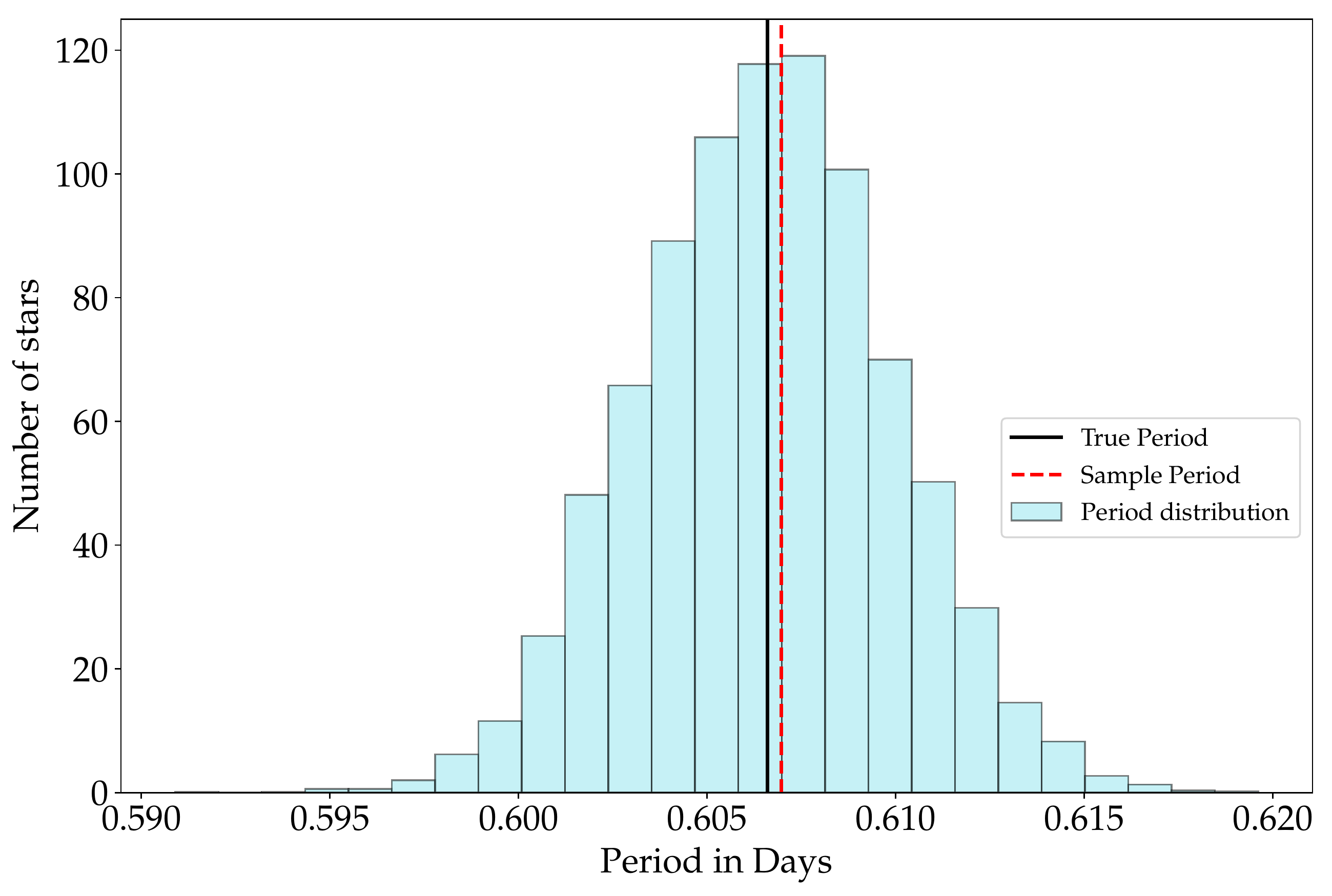}
\caption{For each synthetic LC, a period value (red vertical line) is randomly sampled from a normal distribution, with mean $\textrm{T}$ being the true period of the real LC and $\sigma_{\textrm{T}}$ being the computed uncertainty of the period, $\textrm{T}$.\label{fig:Sampling_new_period}}
\end{figure}

\subsection{Stage 2: Level-wise Data Augmentation in HC}
Since the training set is still imbalanced after aggregating sub-classes into super-classes, we use the three data augmentation techniques described in \S\ref{subsec:data_augmentation}. Each technique is applied and tested independently in our HC based ML pipeline. For the \texttt{SMOTE} approach, features (the mean magnitude, standard deviation, skewness, kurtosis, mean-variance, amplitude and period) described in \citetalias{Hosenie_2019} are extracted from the real LCs. Then, \texttt{SMOTE} automatically balances the class distribution via the creation of synthetic examples sampled over the feature space, such that the size of the minority class equals the size of the majority class, cancelling the imbalance out. For example, considering level 1 in Fig  \ref{fig:Class-distribution-of-full-dataset}, the majority class is Pulsating, consisting of 7338 examples. Therefore, \texttt{SMOTE} adds new examples of the other two minority classes (eclipsing 6312 and rotational 2545) ensuring they both contain 7338 examples. This process is repeated for each branch and level in the HC, where the training set is directly balanced according to the size of the majority class prior to data augmentation.

While for the \texttt{GpFit} and \texttt{RASLE} cases, we are generating new light curves based on real LCs, thus generating new synthetic LC examples. Therefore, our training set will consist of both real and synthetic LCs, whilst we test our ML pipeline with only real LCs. These two techniques can be used to over-sample both the majority and minority class. The number of training examples after augmentation, \textit{A.Tr} used for each level is given in Fig \ref{fig:Class-distribution-of-full-dataset}. Afterwards, features are extracted from these LCs as discussed below.

\subsection{Stage 3: Feature Extraction}
In this work, similarly to \citetalias{Hosenie_2019}, our features are based on 6 intrinsic statistical properties relating to location (mean magnitude), scale (standard deviation), variability (mean variance), morphology (skew, kurtosis, amplitude), and time (period). These features are highly interpretable, and robust against bias \citep{Hosenie_2019}. For the \texttt{GpFit} and \texttt{RASLE} approach, the first six features are extracted directly from the augmented training set (containing both real and synthetic LCs) using the FATS library \citep{Nun_2015}. Whilst for the period feature, the real LCs in the training set are assigned their respective period from the ascii-catalogue \citep{Drake_2017} and the synthetic LCs are assigned a period calculated by the method discussed in \S \ref{subsubsec:period-finding}. For the test set we use only real LCs, hence the six features are extracted directly from the LCs and their period is obtained from the data catalogue. Therefore, we have 7 features that describe each variable star. Fig \ref{fig:real_vs_aug} shows the distribution of the two most important features as investigated in \citetalias{Hosenie_2019} (period and skew) for real and synthetic LCs. We observe that the synthetic LCs show similar characteristics compared to the real LCs.

\subsubsection{Period for augmented LCs}\label{subsubsec:period-finding}
A synthetic LC is given a period based on the uncertainty in the estimated period of the real LC. In this case, the estimated period, $\textrm{T}$, is obtained from \citet{Drake_2017}. The associated uncertainty, $\sigma_{\textrm{T}}$ for a given period is calculated as follows. A periodic signal is detected in a periodogram by the presence of a peak with a certain width and height. In Fourier perspective, we assume that there is a direct relationship between the precision with which a peak's frequency can be detected and the width of this peak; often known as the half-width at half-maximum \citep{vanderplas2018understanding} and is given by:

\begin{equation}
\upsilon_{\frac{1}{2}}\approx\frac{1}{\textrm{{T}}}
\end{equation}

\noindent This can be viewed as interpreting the periodogram with the least-square method, that is, the inverse of the curvature of the peak is determined with the uncertainty \citep{ishak2017statistics}. In the Bayesian perspective, this translates to a Gaussian curve fit to the exponentiated peak \citep{smith2012maximum, bretthorst2013bayesian}. Let us consider a periodogram with maximum value $\textrm{P}_{max}=\textrm{P}\left(\upsilon_{max}\right)$, such that

\begin{equation}
\textrm{P}\left(\upsilon_{max}\pm \upsilon_{\frac{1}{2}}\right)=\frac{\textrm{P}_{max}}{2}.
\end{equation}

\noindent Hence, the Bayesian uncertainty is calculated by approximating the exponentiated peak as a Gaussian, that is,

\begin{equation}
\exp\left[\textrm{P}\left(\upsilon_{max}\pm\delta \upsilon\right)\right]\propto\exp\left[\frac{-\delta \upsilon^{2}}{\left(2\sigma_{\upsilon}^{2}\right)}\right].
\end{equation}

\noindent The above equation can then be written as follows and we obtain the uncertainty in frequency in Eq \ref{eq:uncertainty_frequency}.

\[
\frac{\textrm{P}_{max}}{2}\approx\textrm{P}_{max}-\frac{\upsilon_{\frac{1}{2}}^{2}}{\left(2\sigma_{\upsilon}^{2}\right)};\quad\quad\frac{\upsilon_{\frac{1}{2}}^{2}}{2\sigma_{\upsilon}^{2}}\approx\frac{\textrm{P}_{max}}{2};
\]

\begin{equation}\label{eq:uncertainty_frequency}
\sigma_{\upsilon}\approx\frac{\upsilon_{\frac{1}{2}}}{\sqrt{\textrm{P}_{max}}},
\end{equation}

\noindent where $\delta \upsilon \approx \upsilon_{\frac{1}{2}}$. Considering the signal-to-noise ratio $\varphi=\textrm{rms}\left[\frac{y_{n}-\mu}{\sigma_{n}}\right]$, where $\mu$ is the mean magnitude, $y_{n}$ and $\sigma_{n}$ is the magnitude and error in magnitude for each data point respectively. We can then write the following equation for a well-fitted model,

\begin{equation}\label{eq:Pmax}
\textrm{P}_{max}\approx\frac{\varphi^{2}N}{2}.
\end{equation}

\noindent We then substitute Eq. \ref{eq:Pmax} in Eq. \ref{eq:uncertainty_frequency} and the uncertainty in frequency can be written as:

\begin{equation}
\sigma_{\upsilon}\approx \upsilon_{\frac{1}{2}}\sqrt{\frac{2}{\varphi^{2}N}},
\end{equation}

\noindent where $\upsilon_{\frac{1}{2}}\approx\frac{1}{\textrm{T}},$ $N$ is the number of data points and $\varphi$ is the signal to noise. We now compute the uncertainty in period by taking the derivative of $\sigma_{\upsilon}$,

\[
\frac{\partial \upsilon}{\partial \textrm{T}}\approx-\frac{1}{\textrm{T}^{2}};\quad\quad\partial\textrm{T}=-\textrm{T}^{2}\sigma_{\upsilon};\quad\quad
\sigma_{\textrm{T}}^{2}=\textrm{T}^{4}\sigma_{\upsilon}^{2}.
\]

\noindent Hence, the uncertainty in period is then obtained using Eq \ref{eq:uncertainty_in_period}.

\begin{equation}\label{eq:uncertainty_in_period}
\sigma_{\textrm{T}}=\textrm{T}^{2}\sigma_{\upsilon}
\end{equation}

\noindent where $\sigma_{\textrm{T}}$ will be Gaussian if $\sigma_{\upsilon}$ is very small. A period value is given to each synthetic LC (generated either with \texttt{GpFit} or \texttt{RASLE}), by randomly sampling from a normal distribution with mean, $\textrm{T}$ (the true period of the real LC from which the synthetic LCs are generated) and within 1 $\sigma$-confidence interval, being $\sigma_{\textrm{T}}$ using Eq. \ref{eq:uncertainty_in_period}. An example of associating a period to an augmented LC is shown in Fig \ref{fig:Sampling_new_period}.

\subsection{Stage 4: Training with Bayesian Optimization}\label{Training with Bayesian Optimization}
We first randomly split our data into training (70 per cent) and testing sets (30 per cent). The training set moves through the first level in the HC scheme discussed in \S\ref{subsec:HC_Scheme}. The training examples are then augmented using one of the three data augmentation techniques and features are extracted where appropriate. Afterwards, the model (see dotted square at level 1 in Fig \ref{fig:Class-distribution-of-full-dataset}) is trained using either the RF or XGBoost classifier, as required. We then use a Bayesian Optimization approach to find the best hyper-parameters for the ML algorithm. It has been demonstrated for large parameter spaces that Bayesian Optimization, also known as Sequential Model-Based Optimization (SMBO, \citealt{hutter2011sequential}) performs better than either manual or randomized grid searches \citep{bergstra2013hyperopt}. It is one of the most efficient techniques for hyper-parameter optimization of ML algorithms.

In this work, we used SMBO techniques compared to \citetalias{Hosenie_2019}, who used a randomized grid-search for hyper-parameter optimization. Before applying the above methods, we perform a stratified cross validation. The training data is split into 5 folds, where 4 different folds are kept for training each time and an independent fold is used for validation. We then use the SMBO method, \texttt{HyperOpt} \citep{bergstra2013hyperopt} to find the best hyper-parameters on the 4 folds and validated the model on the independent fold. We then evaluate our trained model based on balanced-accuracy, G-mean, precision, recall, and F1-scores, on real LCs in the test set.

\section{Analysis and Results}\label{sec:Analysis_and_Results}

This paper is mostly concerned with learning from an imbalanced class distribution. The problem is typically addressed using the following approaches.
\begin{enumerate}
\item \noindent \textit{Data level}: We employ three approaches to the HC scheme in such a way that the class distributions are rebalanced directly; that is, it is a first proof of principle application of a level-wise augmentation in Hierarchical taxonomy, where we resample the original dataset to achieve a desired balancing.
\item \noindent \textit{Algorithm level}: We focus on using two different algorithms (RF and XGBoost), together with a Bayesian Optimization algorithm for hyper-parameter tuning, to achieve improved performance on the minority class examples.
\end{enumerate}

\begin{figure}
\centering
\includegraphics[width=0.5\textwidth]{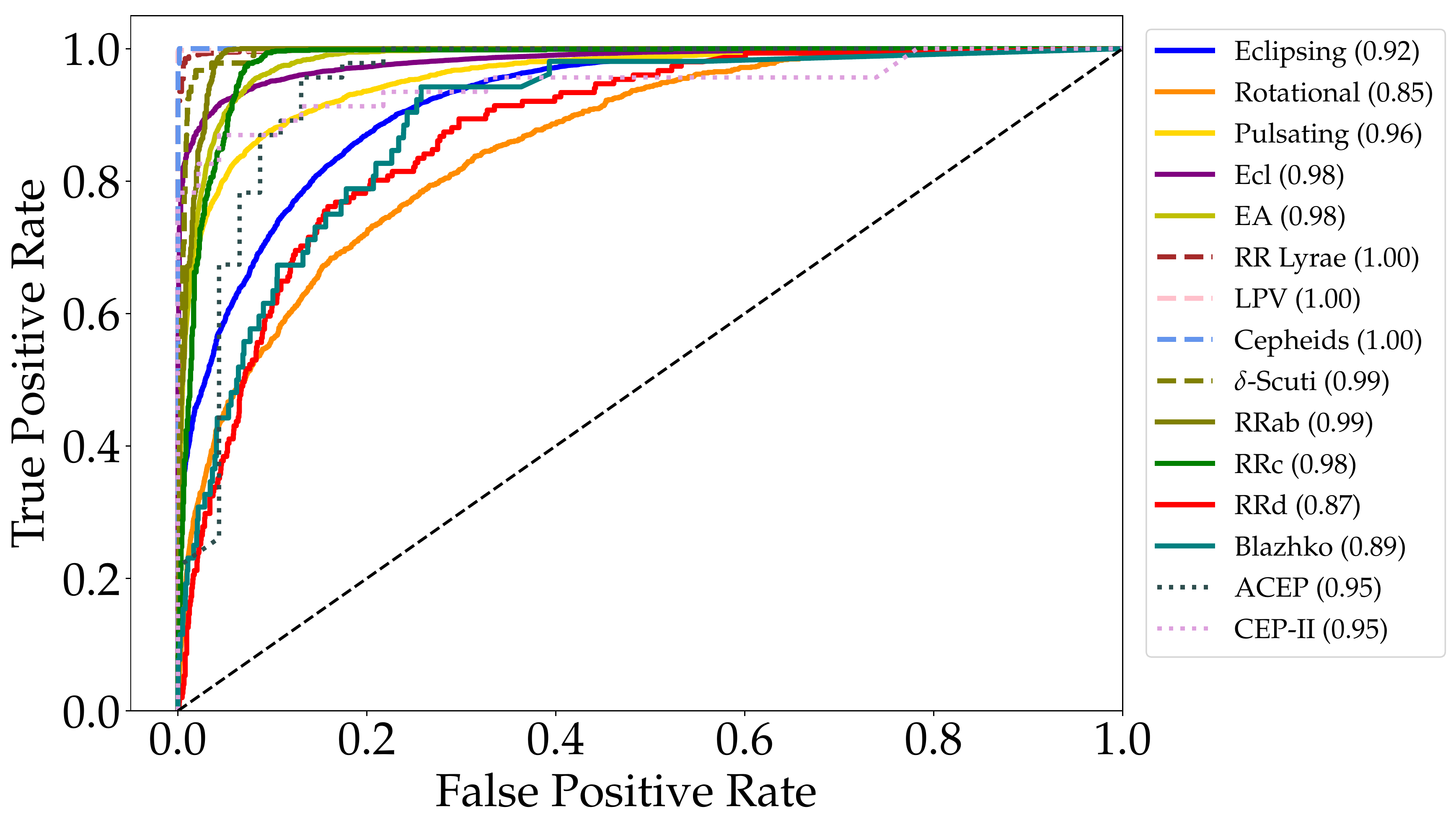}
\caption{Receiver operating characteristic (ROC) curves for each node in the hierarchical model. Each curve represents a different variable star class with the area under the ROC curve (AUC) score in brackets. This metric is computed on the 30\% of the dataset used for testing.\label{fig:ROC_cuves}}
\end{figure}

The HC algorithm is trained on both real and artificially augmented data and tested on real data. We show the results of the three data augmentation techniques in Table \ref{tab:HC_results_data_augmentation}. We assess the consistency of the results based on balanced-accuracy and G-mean scores. The shaded blue color represents the augmentation methods, which when applied together with the HC classifier, yielded improved results over \citetalias{Hosenie_2019}. We found that \texttt{GpFit} achieves the best performance measures compared to \citetalias{Hosenie_2019} at all levels in the HC. When using the \texttt{GpFit} method, we found that our RF implementation performs best at all HC levels when compared to \citetalias{Hosenie_2019} and we highlight this result in gray. In addition, we found that XGBoost, similarly to the RF, provides good performance for variable star classification. Moreover, in \citetalias{Hosenie_2019}, we show that the HC model is neither underfitting nor overfitting by plotting precision-recall curves at different levels. In this paper, we assess the consistency of the results using \texttt{GpFit} and RF by plotting the Receiver Operator Characteristic (ROC) curve for each class (see Fig \ref{fig:ROC_cuves}). We note that classification performance is very good. The area under the ROC curve (AUC) values are greater than 0.95 for several classes, except for Rotational, RRd, and Blazhko. The reasons for these misclassification are further discussed in \S \ref{subsec:imbalance_reason}.

We improve upon the result obtained in \citetalias{Hosenie_2019}. For instance, the balanced-accuracy increases from 61 to 65 per cent in level 1, from 86 to 88 per cent at level 2 for the eclipsing node, from 86 to 87 per cent for sub-classes of RR Lyrae at level 3, and finally from 81 to 83 per cent for Cepheids at level 3. To check the consistency and robustness of our new approach, we perform an additional step. We use different splits ($K$ = 5, 6, \ldots,10) during cross-validation and predict on the 30\% test set. With these analyses, we obtain an uncertainty on the metric scores considered, for example for Cepheids at level 3, a 0.83 $\pm$ 0.02 balanced-accuracy and  0.91 $\pm$ 0.01 G-mean score are obtained. We obtain similar results at different levels in the hierarchy. In these analyses, we observe that we have not made a huge improvement to \citetalias{Hosenie_2019}, in terms of minority classes and we explain the various reasons that might lead to this outcome in \S \ref{subsec:imbalance_reason}.

\renewcommand{\arraystretch}{1.4}
\begin{table*}
\begin{minipage}{150mm}

\caption{Evaluation metrics used to summarize the HC pipeline with the application of three methods of data augmentation. We present the balanced-accuracy and G-mean scores level-wise to evaluate our model. \citetalias{Hosenie_2019} results are presented in bold text in the table. It is seen that the HC pipeline performs fairly well with data augmentation, achieving G-mean scores above $\sim$ 80\% at all levels. The shaded blue represents the augmentation methods that outperform \citetalias{Hosenie_2019}. We observe that at all levels, \texttt{GpFit} together with RF, performs better than \citetalias{Hosenie_2019} and it is represented in shaded gray. The `$\sim$' represents a single value for the computed average metrics by taking into consideration the overall classes.}
\label{tab:HC_results_data_augmentation}
\noindent \begin{centering}
\begin{tabular}{|c|c|c|c|}
\hline 
\textbf{\textcolor{black}{Augmentation Techniques}} & \textbf{\textcolor{black}{Classifiers}} & \textbf{\textcolor{black}{G-Mean}} & \textbf{\textcolor{black}{Balanced-accuracy}}\tabularnewline
\hline 
\multicolumn{4}{|c|}{\textcolor{blue}{First Level: Eclipsing, Rotational and Pulsating Classification}}\tabularnewline
\hline 
{\citetalias{Hosenie_2019} (No augmentation)} & {\textbf{RF}} & {\textbf{0.78/0.78/0.86 ($\sim$ 0.79)}}  & {\textbf{0.59/0.60/0.75 ($\sim$ 0.61)}}\\
\hline 
\cellcolor{LightCyan} & XGBoost & 0.80/0.77/0.89 ($\sim$ 0.81) & 0.63/0.59/0.80 ($\sim$ 0.65)\tabularnewline
\cellcolor{LightCyan}\multirow{-2}{1cm}{SMOTE} & RF  & 0.80/0.78/0.89 ($\sim$ 0.81)  & 0.63/0.60/0.79 ($\sim$ 0.65) \tabularnewline
\hline 
\cellcolor{LightCyan} & XGBoost & 0.82/0.76/0.89 ($\sim$ 0.83) & 0.66/0.57/0.79 ($\sim$ 0.68)\tabularnewline
\cellcolor{LightCyan}\multirow{-2}{1cm}{RASLE} & RF & 0.82/0.77/0.89 ($\sim$ 0.83) & 0.66/0.58/0.79 ($\sim$ 0.68)\tabularnewline
\hline 
\cellcolor{LightCyan} & XGBoost & 0.80/0.75/0.89 ($\sim$ 0.81) & 0.63/0.56/0.79 ($\sim$ 0.65)\tabularnewline
\cellcolor{LightCyan}\multirow{-2}{1cm}{GpFit} & \cellcolor{Gray}{RF} & \cellcolor{Gray}{0.80/0.75/0.89 ($\sim$ 0.81)} & \cellcolor{Gray}{0.63/0.56/0.78 ($\sim$ 0.65)}\tabularnewline
\hline 
\multicolumn{4}{|c|}{\textcolor{blue}{Second Level: RR Lyrae, LPV, Cepheids and $\delta$-Scuti}}\tabularnewline
\hline 
{\citetalias{Hosenie_2019} (No augmentation)} & {\textbf{RF}} & {\textbf{0.99/1.00/0.97/1.00 ($\sim$ 0.99)}} & {\textbf{0.98/0.99/0.93/1.00 ($\sim$ 0.98)}}\\
\hline 
\multirow{2}{*}{SMOTE} & XGBoost & 0.99/1.00/1.00/0.95 ($\sim$ 0.99) & 0.97/0.99/1.00/0.90 ($\sim$ 0.97)\tabularnewline
 & RF & 0.99/1.00/1.00/0.96 ($\sim$ 0.99) & 0.97/0.99/1.00/0.92 ($\sim$ 0.97)\tabularnewline
\hline 
\multicolumn{1}{|c|}{\cellcolor{LightCyan}} & XGBoost & 0.99/1.00/0.99/0.93 ($\sim$ 0.99) & 0.98/1.00/0.98/0.85 ($\sim$ 0.98)\tabularnewline
\cellcolor{LightCyan}\multirow{-2}{1cm}{RASLE} & RF & 0.99/1.00/1.00/0.94 ($\sim$ 0.99) & 0.98/0.98/1.00/0.88 ($\sim$ 0.98)\tabularnewline
\hline 
\multicolumn{1}{|c|}{\cellcolor{LightCyan}} & XGBoost & 0.99/1.00/0.99/0.95 ($\sim$ 0.99) & 0.97/0.99/0.97/0.99 ($\sim$ 0.98)\tabularnewline
\cellcolor{LightCyan}\multirow{-2}{1cm}{GpFit} & \cellcolor{Gray}{RF} & \cellcolor{Gray}{0.99/1.00/1.00/0.97 ($\sim$ 0.99)} & \cellcolor{Gray}{0.97/0.99/1.00/0.93 ($\sim$ 0.98)}\tabularnewline
\hline 
\multicolumn{4}{|c|}{\textcolor{blue}{Second Level: Ecl and EA}}\tabularnewline
\hline 
{\citetalias{Hosenie_2019} (No augmentation)} & {\textbf{RF}} & {\textbf{0.93/0.93 ($\sim$ 0.93)}} & {\textbf{0.86/0.86 ($\sim$ 0.86)}}\\
\hline 
\multicolumn{1}{|c|}{\cellcolor{LightCyan}} & XGBoost & 0.94/0.94 ($\sim$ 0.94) & 0.88/0.88 ($\sim$ 0.88)\tabularnewline
\cellcolor{LightCyan}\multirow{-2}{1cm}{SMOTE} & RF & 0.94/0.94 ($\sim$ 0.94) & 0.88/0.88 ($\sim$ 0.88)\tabularnewline
\hline 
\multirow{2}{*}{RASLE} & XGBoost & 0.93/0.93 ($\sim$ 0.93) & 0.85/0.85 ($\sim$ 0.85)\tabularnewline
 & RF & 0.93/0.93 ($\sim$ 0.93) & 0.85/0.86 ($\sim$ 0.86)\tabularnewline
\hline 
\multicolumn{1}{|c|}{\cellcolor{LightCyan}} & XGBoost & 0.93/0.93 ($\sim$ 0.93) & 0.88/0.88 ($\sim$ 0.88)\tabularnewline
\cellcolor{LightCyan}\multirow{-2}{1cm}{GpFit} & \cellcolor{Gray}{RF} & \cellcolor{Gray}{0.94/0.94 ($\sim$ 0.94)} & \cellcolor{Gray}{0.87/0.88 ($\sim$ 0.88)}\tabularnewline
\hline 
\multicolumn{4}{|c|}{\textcolor{blue}{Third Level: RRab, RRc, RRd and Blazhko}}\tabularnewline
\hline 
{\citetalias{Hosenie_2019} (No augmentation)} & {\textbf{RF}} & {\textbf{0.97/0.92/0.65/0.44 ($\sim$ 0.92)}} & {\textbf{0.94/0.85/0.40/0.18 ($\sim$ 0.86)}}\\
\hline 
\multirow{2}{*}{SMOTE} & XGBoost & 0.95/0.92/0.67/0.58 ($\sim$ 0.91) & 0.91/0.83/0.42/0.31 ($\sim$ 0.83)\tabularnewline
 & RF & 0.95/0.82/0.47/0.33 ($\sim$ 0.91) & 0.91/0.82/0.47/0.33 ($\sim$ 0.83)\tabularnewline
\hline 
\multicolumn{1}{|c|}{\cellcolor{LightCyan}} & XGBoost & 0.96/0.95/0.56/0.53 ($\sim$ 0.92) & 0.93/0.89/0.30/0.26 ($\sim$ 0.87)\tabularnewline
\cellcolor{LightCyan}\multirow{-2}{1cm}{RASLE} & RF & 0.97/0.95/0.52/0.52 ($\sim$ 0.92) & 0.94/0.90/0.25/0.25 ($\sim$ 0.87)\tabularnewline
\hline 
\multicolumn{1}{|c|}{\cellcolor{LightCyan}} & XGBoost & 0.97/0.93/0.57/0.44 ($\sim$ 0.92) & 0.94/0.86/0.30/0.17 ($\sim$ 0.85)\tabularnewline
\cellcolor{LightCyan}\multirow{-2}{1cm}{GpFit} & \cellcolor{Gray}{RF} & \cellcolor{Gray}{0.97/0.93/0.56/0.41 ($\sim$ 0.92)} & \cellcolor{Gray}{0.94/0.87/0.32/0.26 ($\sim$ 0.87)}\tabularnewline
\hline 
\multicolumn{4}{|c|}{\textcolor{blue}{Third Level: ACEP and Cep-II}}\tabularnewline
\hline 
{\citetalias{Hosenie_2019} (No augmentation)} & {\textbf{RF}} & {\textbf{0.90/0.90 ($\sim$ 0.90)}} & {\textbf{0.82/0.80 ($\sim$ 0.81)}}\\
\hline 
\multirow{2}{*}{SMOTE} & XGBoost & 0.88/0.88 ($\sim$ 0.88) & 0.78/0.76 ($\sim$ 0.77)\tabularnewline
 & RF & 0.88/0.88 ($\sim$ 0.88) & 0.78/0.76 ($\sim$ 0.77)\tabularnewline
\hline 
\multicolumn{1}{|c|}{\cellcolor{LightCyan}} & XGBoost & 0.88/0.88 ($\sim$ 0.88) & 0.77/0.78 ($\sim$ 0.77)\tabularnewline
\cellcolor{LightCyan}\multirow{-2}{1cm}{RASLE} & RF & 0.88/0.88 ($\sim$ 0.88) & 0.77/0.78 ($\sim$ 0.78)\tabularnewline
\hline 
\multicolumn{1}{|c|}{\cellcolor{LightCyan}} & XGBoost & 0.88/0.88 ($\sim$ 0.88) & 0.78/0.78 ($\sim$ 0.78)\tabularnewline
\cellcolor{LightCyan}\multirow{-2}{1cm}{GpFit} & \cellcolor{Gray}{RF} & \cellcolor{Gray}{0.91/0.91 ($\sim$ 0.91)} & \cellcolor{Gray}{0.84/0.82 ($\sim$ 0.83)}\tabularnewline
\hline 
\end{tabular}
\par\end{centering}
\end{minipage}
\end{table*}

\subsection{Impact of imbalance on classification performance}\label{subsec:imbalance_reason}
\begin{figure}
\centering
\subfloat[]{\includegraphics[width=0.17\textwidth]{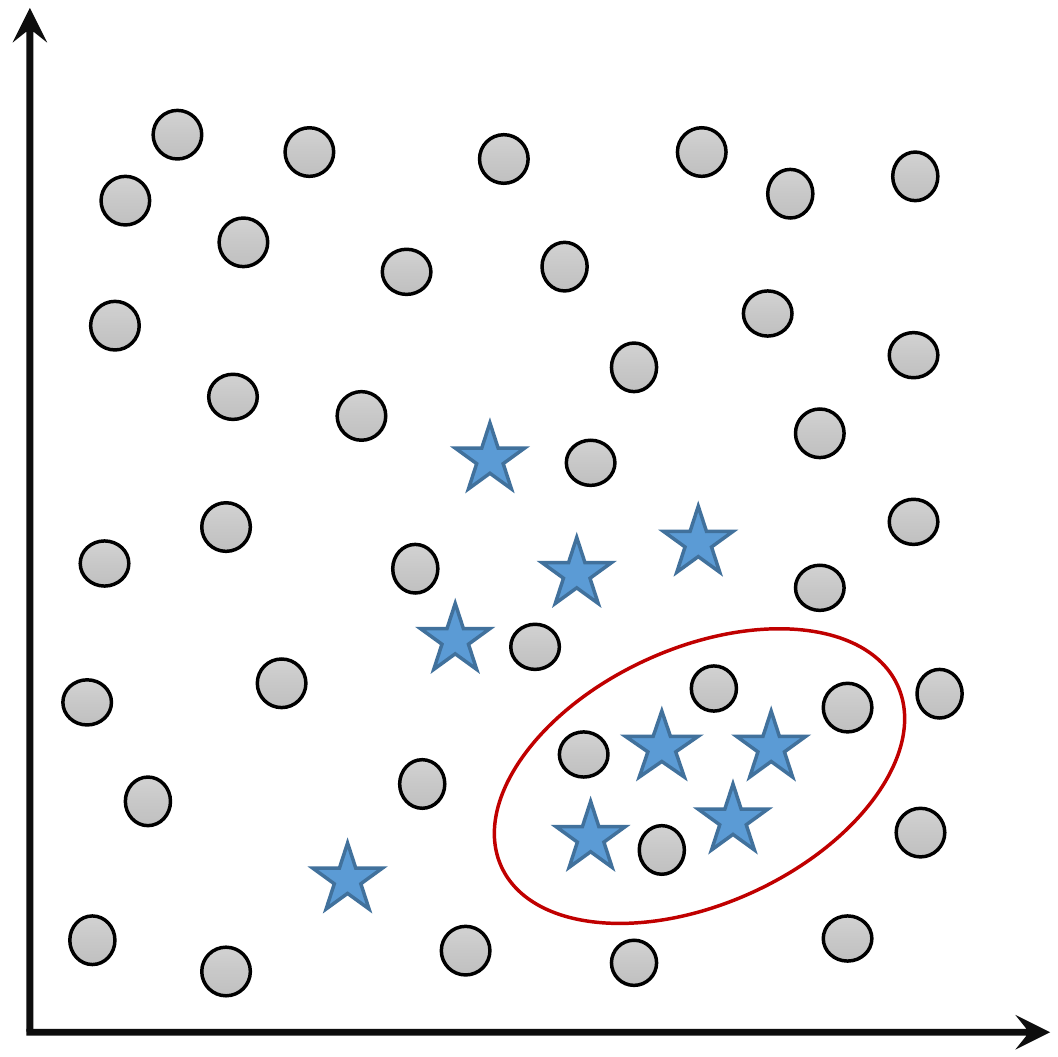}}
\subfloat[]{\includegraphics[width=0.17\textwidth]{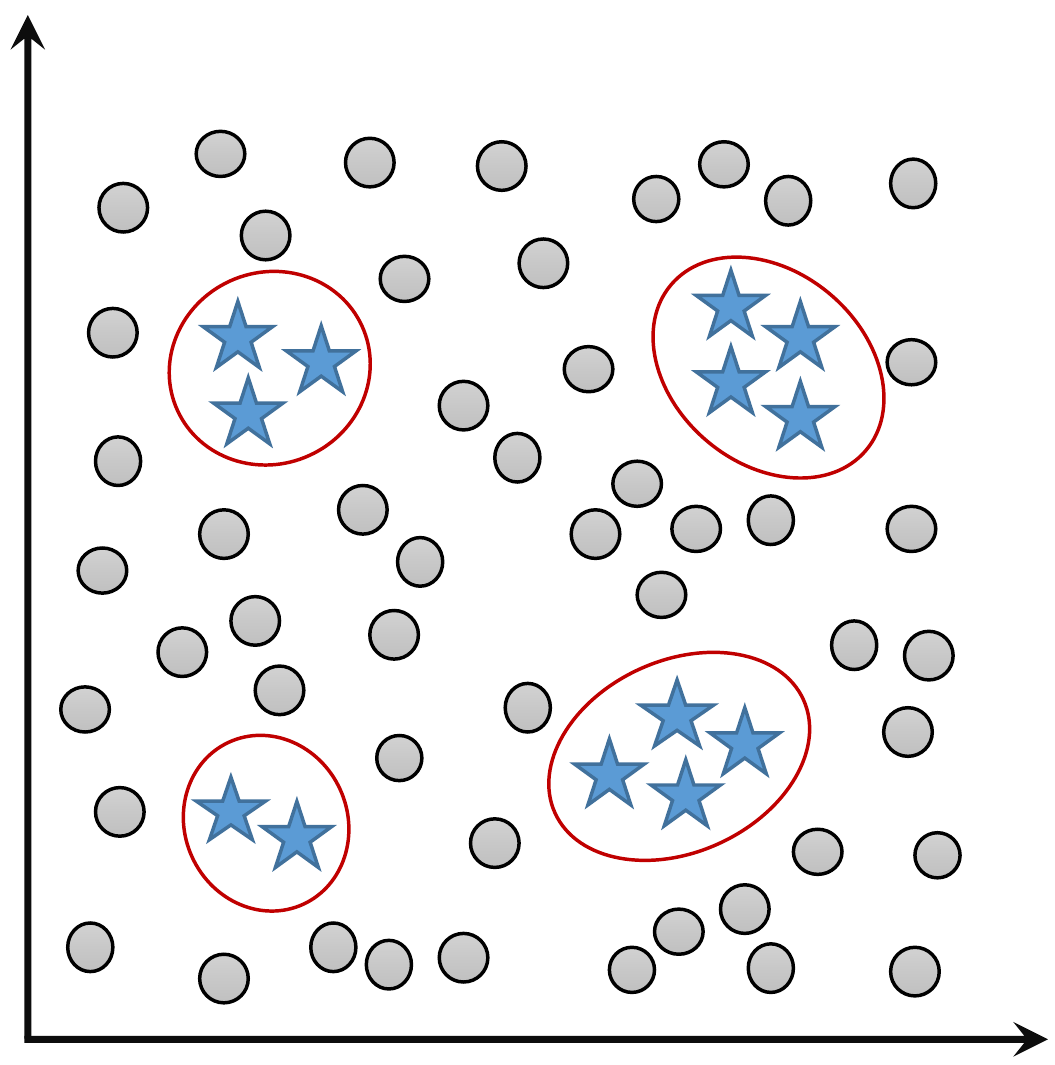}}
\caption{\label{fig:class_inseparability_small_disjuncts}Demonstrattion of (a) Class inseparability and (b) small disjuncts in feature space.}
\end{figure}

\begin{figure*}
\centering
\subfloat[]{\includegraphics[width=0.50\textwidth]{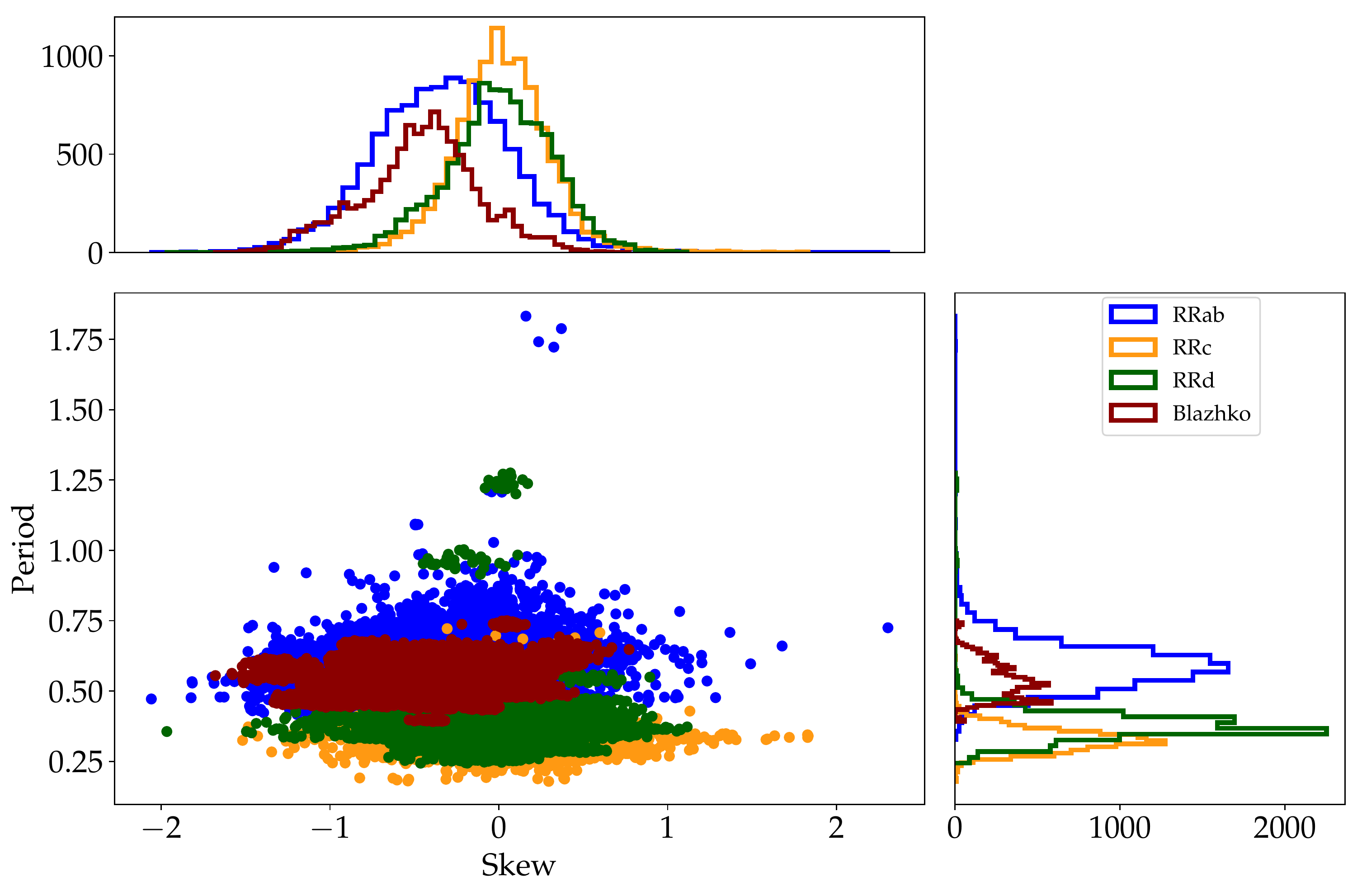}}
\subfloat[]{\includegraphics[width=0.50\textwidth]{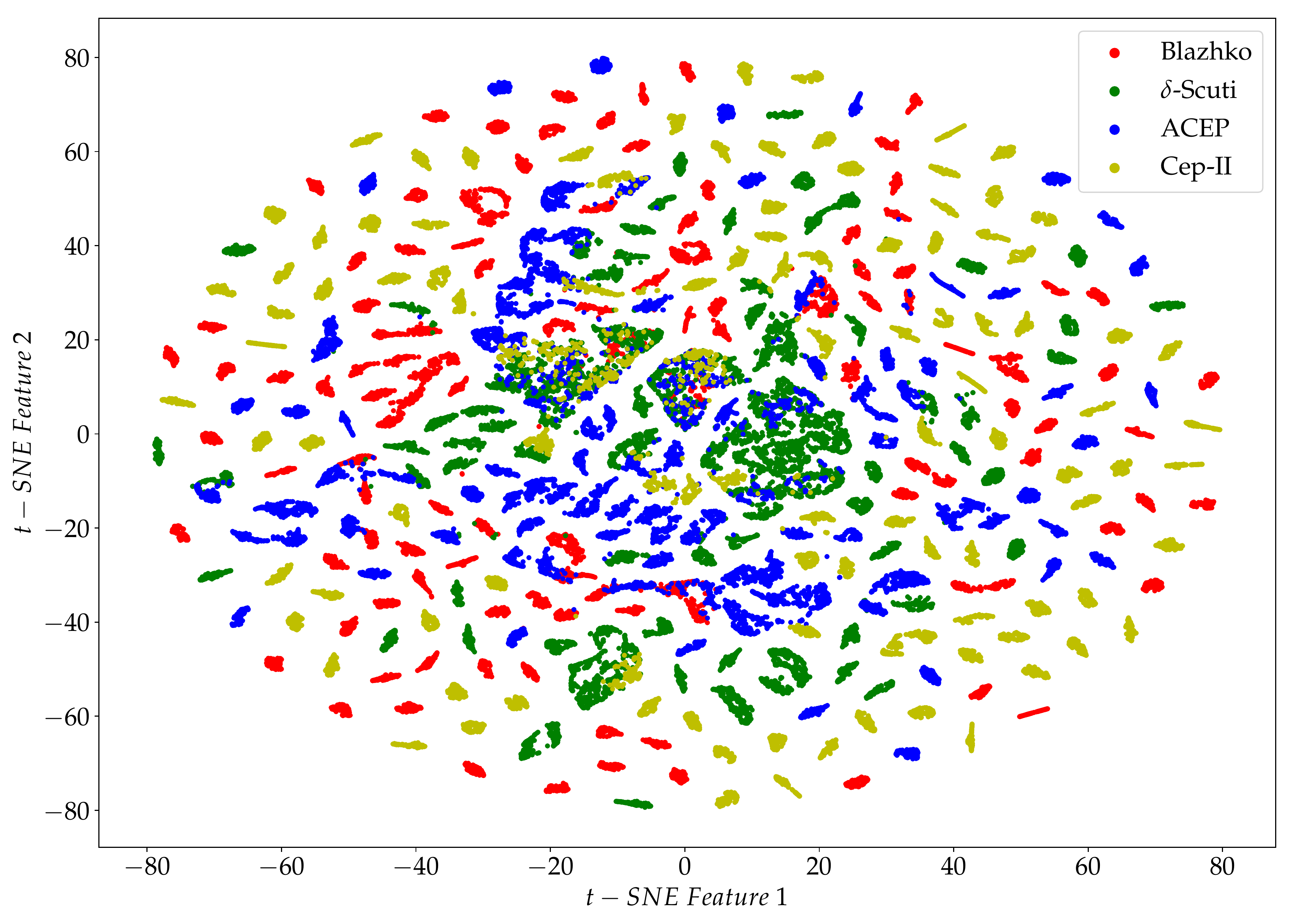}}
\caption{\label{fig:RR_Lyrae_tsne}(a) shows the Period-Skew distribution of RRab, RRc, RRd and Blazhko after augmenting each respective class to 10,000 examples. We note that the classes are still overlapping in the feature space, even after the augmentation process. (b) illustrates small disjoints in feature space using t-distributed stochastic neighbour embedding (t-SNE) visualization in the small sample size data (Blazhko, $\delta$-Scuti, ACEP and Cep-II), after augmentation. No distinct separation is seen within the feature space.}
\end{figure*}

Training a classifier upon imbalanced data does not guarantee poor generalisation performance \citep{galar2011review}. Regardless of imbalance, if the features or the training data themselves are discriminative enough to provide a clear separation between the different classes, then classifiers will likely generalize well. However, there are three main characteristics of imbalanced data sets that make it hard for a classifier to discriminate the minority from the majority classes. These are
\begin{enumerate}
\item \noindent small sample sizes \citep{galar2011review,he2008learning},
\item \noindent class inseparability \citep{galar2011review, japkowicz2002class} (see Fig \ref{fig:class_inseparability_small_disjuncts}(a) \& \ref{fig:RR_Lyrae_tsne}) and,
\item \noindent small disjuncts (see Fig \ref{fig:class_inseparability_small_disjuncts}(b)).
\end{enumerate}

Ultimately, the training data showing these characteristics conspire to make it hard for any classifier to build an optimal decision boundary leading to sub-optimal classifier performance. These characteristics are seen at some levels in the HC. In this section, we illustrate these effects at level 3 using the sub-classes of RR Lyrae. Fig \ref{fig:class_inseparability_small_disjuncts}(a) shows that some classes have overlapping characteristics, which leads to poor performance. We observe similar characteristics (class-overlapping) for the sub-classes of RR Lyrae in Fig \ref{fig:RR_Lyrae_tsne}(a), even after balancing the classes in the training set. These overlapping characteristics are due to the fact that there are no physical distinction between some of the subclasses. As can be seen in Fig \ref{fig:RR_Lyrae_tsne}(a), RRab and RRc classes can reasonably be separated based on their period alone. RRab are variable stars pulsating in fundamental mode, RRc stars pulsate in the first overtone while RRd stars simultaneously pulsate in the fundamental and first overtone. Therefore, RRd's form part of both RRab and RRc variable stars at the same time. In addition, Blazhko stars are found among RRab stars \citep{jurcsik2009konkoly}, RRc stars \citep{netzel2018blazhko} and even RRd stars \citep{jurcsik2015overtone}. This explains the poor performance of the classifier for separating RRd and Blazhko stars, even after balancing the classes. In addition, we also present a t-distributed stochastic neighbour embedding (t-SNE, \citealt{vanderMaaten_2008}) of the minority classes (Blazhko, $\delta$-Scuti, ACEP \& Cep-II) in Fig \ref{fig:RR_Lyrae_tsne}(b) after augmenting them using the \texttt{GpFit} method. The result shown in Fig \ref{fig:RR_Lyrae_tsne}(a) does not differ when we perform multiple runs with different parameters. Each time we find small disjuncts in the feature space, showing characteristics similar to those shown in Fig \ref{fig:class_inseparability_small_disjuncts}(b), thus making it difficult for the classifier to construct a decision boundary. 

In this paper, we found that training the HC with class-balanced data has the effect of improving balanced-accuracy and G-mean scores. However, the minority classes are still misclassified. Although these results suggest that balancing the class distribution is not sufficient for classifying the minority classes, their capacity to prevent overfitting and increase the recall rate makes them appealing.

Another reason that leads to misclassification - the lack of a standard set of correctly classified (i.e. where the ground truth is certain) variable star example useful for training. \citet{Drake_2017} investigated the level of agreement of their classifications with the International Variable Star Index (VSX, \citet{watson2006international}). They found that 

\begin{enumerate}
\item \noindent VSX has not classified any of their Blazhko stars, but instead simply classify them as RRab stars,
\item \noindent VSX classified many of their contact binaries as detached and semi-detached binaries,
\item \noindent most of their Rotational stars (spotted or ellipsoidal variables) have been classified as contact binaries, and
\item \noindent most of their RRd stars have been misclassified as other stars (RRab, RRc) by VSX.
\end{enumerate}

We observe similar misclassifications when using our automated HC pipeline, even after balancing the classes. With the presence of so many misclassified objects, we can plausibly say neither \citet{Drake_2017} or VSX can be considered as providing ground truth. Therefore, there is a real need to have a standard set of correctly identified variable stars that can be utilized for training automated machine learning methods. It is imperative to train these sophisticated ML based algorithms with accurately calibrated priors in order to obtain reliable classification outputs.

\section{Conclusion}\label{sec:Conclusion_2}
In this paper, we present a new approach for tackling the problem of imbalanced data: a level-wise data augmentation in a hierarchical classification framework. Through an empirical investigation, we demonstrate three techniques for augmenting data, that is, \texttt{SMOTE}, \texttt{RASLE} and \texttt{GpFit} are applied to variable star data. We show that using RF and \texttt{GpFit} together can effectively improve recall rates, hence increasing the balanced-accuracy and G-mean scores by 1-4 per cent. Although, the results show that even after balancing the training set level-wise, such approaches do not prevent the misclassification of the minority class, though their capacity to increase other metrics (e.g. recall) still makes their application appealing. Perhaps, the misclassification occurs because these objects are just not easily separable and we observe similar misclassifications in this paper as determined by \citet{Drake_2017} when they compared their results with VSX. Therefore, it is imperative to have correctly labelled data that can accurately be used to train automated ML pipeline in order to output reliable classification performance.

\section*{Acknowledgements}
We thank the referee for useful comments and suggestions for improving this paper. ZH acknowledges support from the UK Newton Fund as part of the Development in Africa with Radio Astronomy (DARA) Big Data project delivered via the Science \& Technology Facilities Council (STFC). BWS acknowledges funding from the European Research Council (ERC) under the European Union's Horizon 2020 research and innovation programme (grant agreement No. 694745). AM is supported by the Imperial President's PhD Scholarship. VMB acknowledges funding from the National Research Foundation of South Africa (grant numbers 98969 and 119446).

\bibliographystyle{mnras}
\bibliography{../Project-1-Variable-stars-classification/reference.bib}

\bsp	
\label{lastpage}
\end{document}